# Potential of Photovoltaics and Energy Storage to Address Lack of Electricity Access

GF L'Her[1†], AG Osborne[1†], AE Schweikert[1], CS Ramstein[2], BL Stoll[3], MR Deinert[1,4]*

[1]Department of Mechanical Engineering, The Colorado School of Mines
[2]The World Bank, Washington, D.C.
[3]Department of Mechanical Engineering, The University of Texas at Austin
[4]Payne Institute for Public Policy, The Colorado School of Mines

**Abstract**. Lack of electricity access is widespread in the developing world and associated with increased mortality, reduced educational levels, and economic and social disadvantages, especially among women. The 2030 Agenda for Sustainable Development has emphasized securing access to affordable, reliable, and sustainable energy for all. For climatological and health reasons, particular attention has been focused on expanding the use of renewables for electricity production. In particular, photovoltaics, coupled to energy storage, is an attractive option for dispatchable electricity production, but the degree to which they can be used to address global lack of electricity access, and associated costs, merits more attention. This study presents a global geospatial analysis to identify areas suitable for production of dispatchable electricity using photovoltaics and energy storage. Analysis considers land use restrictions, 25 years of historical hourly solar irradiance, seasonal demand curves, population, and visible nighttime light (as a measure of electrification). We show that nearly all the population identified without electricity access (approx. 1.1 billion people) could get access to Tier 5 level electricity in the Sustainable Energy for All initiative framework using photovoltaics and battery storage coupled systems. Under most cost scenarios analyzed, around 90% of this population could be served for a lifetime cost of electricity (LCOE) of $0.20/kWhe or less at current system costs.

**\*Corresponding Author.** MR Deinert, Department of Electrical and Computer Engineering, Cornell University. Email: mrd6@cornell.edu.

[†]*The first two authors contributed equally to this work*.

**Introduction**. The link between access to reliable energy sources and economic development is well documented [1]–[5]. Existing targets to electrify all persons globally by 2030 have resulted in substantial and ongoing energy development efforts in regions with low electricity access, notably in Sub-Saharan Africa. However, recent estimates show that over 2.8 billion people globally still rely on biomass for cooking and heating [2], with 1.0 to 1.6 billion people having no access to electricity [6], [7], and the developing world is not on track to meet the seventh sustainable development goals [8]. These disparities are uneven around the globe. Regionally, lack of electricity access is particularly high in Sub-Saharan Africa, where it can exceed 50% of the population [7]. Current estimates still show that around 650 million people may not have access by 2030, the vast majority of whom will be in Sub-Saharan Africa [9]. Even within Sub-Saharan Africa, there are major differences, with countries such as Nigeria home to nearly 90 million people without access to electricity, while countries such as Kenya or Rwanda are on track to reach universal access by 2030 [8], [10]. The goal of expanding energy access, in particular, requires a consideration of climate issues associated with emissions as well as questions of economic feasibility, social and environmental impact [11]. Concerns about electricity sources carbon signature, reliability, and resilience should also be inherent in considerations of expanding global energy access. However, the nexus between climate and development is complex and lacks focus, with different perspectives on the relative costs and benefits affecting how each is prioritized. The Sustainable Development Goals outlined by the United Nations highlight this complexity. Each of these 17 goals – and 65% of the 169 targets - have been shown to not only have ties to energy, but also to require action on the choice and development of energy systems if the goals are to be met [12].

The 2030 Agenda for Sustainable Development highlights the clear ties between energy access, poverty reduction, and need for a low-carbon growth pattern. The ambitious target of energy access for all requires new ways of providing reliable, low-carbon energy. The tension between an urgent need to expand access to areas with lack of electricity access (Sustainable Development Goal 7) and many other goals, including addressing climate change (Sustainable Development Goal 13) and facing its impacts, provide an opportunity to evaluate the feasibility of renewable technologies in this paradigm [10], [12].

Dramatic capital cost reductions in recent years have made photovoltaic panels an attractive option for electricity production without the climate and environmental consequences of fossil-fuel technologies. Between 2009 and 2019, solar photovoltaic generation increased globally from 20TWh to 720TWh [14], while the cost of battery storage has dropped dramatically – nearly 70% between 2015 and 2018 alone [15]. Rapid cost declines coupled with the growing consensus that photovoltaics is a mature technological option make it an ideal technology for many locations, providing economical and scalable carbon-free generation [16]. Achieving high levels of variable renewable energy generation, however, is highly dependent on storage capabilities and costs [17]. This has led to multiple studies that have tried to assess the ability of renewable energy systems and storage to meet 100% electricity needs [18], [19], [10].  Unfortunately, these assessments have relied on limited time horizons for understanding the effect of intermittency, failed to consider land use constraints, and relied on assumptions about technologies that are out of step with what is generally assumed to be realistic [20].  Considerations of the potential impact on global lack of electricity access have also been incomplete.

In 2015, Bhatia and Angelou [21] proposed a multi-tiered framework to classify energy access targets in areas with little to no current electrification. The classifications define Tiers 0-5, with Tier 5 indicating high availability and reliability of electricity throughout day and nighttime uses. Tier "0" indicates no access to electricity, while Tier 1 indicates enough power to charge a phone or use a lightbulb for a few hours per day. An important consideration in these Tiered definitions is not simply the daily or annual amounts of energy needed, but the reliability (capacity factor) and timing of available electricity (daytime compared with night). As access increases along Tiers, larger appliances are included causing electricity demand to fluctuate with the time of day and season of the year. In addition to household uses, the Tiered system covers both community and productive uses, including streetlights.

Much of the literature in this area focuses on least-cost optimization models of clean technologies (e.g., [22]–[25]) in small energy-impoverished regions. Different studies looking at the potential of concentrating solar power, photovoltaics, wind or other resource assessment possibilities in specific geographies include Southern Africa [26]–[29], specific African countries [22], [30] and Europe [31], [32]. One study on concentrating solar power investigates the possibility of baseload generation in desert regions of US and China using load profiles from high-demand areas and high-voltage transmission infrastructure. The study considers irradiance values, siting terrain considerations, and design optimization using (n-1) load curve reliability methods [24]. A 2018 study limited to Sub-Saharan Africa [33] assesses the feasibility of only the Tier 5 Access levels using household solar photovoltaics coupled to battery storage. A 2021 study looked at how diesel and solar systems could compete in current markets to alleviate the lack of electricity access based on modest access (between tier 3 and tier 4) [34]. Daily average solar insolation data is used and up-sampled to achieve hourly resolution at 1-degree geospatial resolution. The analysis uses a 10% discount rate over 20 years to calculate the levelized cost of electricity using photovoltaic cost data from IRENA as well as estimates for battery costs. The study finds that photovoltaics and battery storage may be a viable option in many locations based on the levelized cost computed. A second paper [35] presents a tool for Sub-Saharan Africa and is used in several studies [22], [35]–[37] to explore planning options for wind, photovoltaics, hydropower or other technologies. Another study identifies access needs across Sub-Saharan Africa using the Tiered system [38]. In [35], the authors perform a continent-wide assessment of grid-extension costs and a comparison of technologies through optimization modeling. The Global Electrification Platform also aims at modeling different pathways to reaching announced targets [cite electrifynow.energydata.info].

A critical gap in these analyses is the combination of technical and high-level siting feasibility alongside considerations of required infrastructure, and the resulting costs for different Tiers of electricity access. This is essential because power cannot be provided without suitable siting of systems and dispatch of electricity. Importantly, the size of those systems and related costs also change dramatically with the amount and reliability of power needed.

In this study, a global assessment of the potential for solar photovoltaics and battery storage to meet the electrification needs of people living with little to no access to electricity is performed. The analysis is done using multi-criteria land-use parameters and an electricity dispatch model. The land-use constraints developed for this study (e.g., slopes, terrain classification, population density, etc.) are important considerations for system siting and a useful intermediate output that can be used for multiple purposes [38]. Further, this study assesses the optimized electricity generation for every location in the world up to a latitude of $60^\circ$, with a focus on regions with lack of electricity access. The assessment is based on population energy demand profiles and solar irradiance that are used as inputs to an energy flow model. The model simulates dispatch to meet target capacity factors and other constraints for each relevant Tier (3-5). Tiers 3-5 are assessed for grid scale arrays and battery systems, which is consistent with the description of Tiered systems including centralized delivery and payment structures [21].

Satisfying electricity needs affordably requires systems to be optimized according to location-specific load curves that characterize a population's diurnal and seasonal electricity demand. Recent work has suggested incorporating local demand-use patterns when considering system sizing in regions with low electricity access (e.g. [39]–[41]). Demand data, however, is rarely available and often requires detailed survey data or similar data gathering. Narayan et al. [39] provides estimated load curves for each Tier of analysis, specifically focused on load profiles from DC-based appliances used in newly electrified regions. We use these Tier-specific household load curves developed for Sub-Saharan Africa, including appliance use appropriate for each tier. These data are used as inputs to our models and aggregated with daily estimates for community and productive uses including streetlights, consistent with the Tiered system guidelines (For additional detail see Methodology below and Supplementary Note 1). Discount rates of 5% and 7% are assessed and levelized cost of electricity (LCOE) is calculated using two cost scenarios for Tiers 3-5. The scenarios assume photovoltaic costs of $1/Watt-peak and both 2025 and 2030 projected costs for battery storage. This results in 12 unique optimization scenarios using different costing models for each Tier, costing scenario, and discount rate specific to population needs in each of the identified locations globally. See Supplementary Note 2 for further details.

**Methods.**
*Land-Use and Electricity Access Estimates*
This study combines global electricity access data, land use parameters that constrain utility-scale construction, estimated electricity demand and available solar irradiance in an energy flow model that calculates electricity dispatch. The flow model is used to find photovoltaic array and battery storage sizes that meet a population's electricity needs according to each Tier of access. For electricity access and siting considerations, satellite imagery and global geospatial data were used (Table 1). Global data for wetland locations were taken from [42], [43] and for protected areas from [44]. Elevation data at 3 arc second resolution were taken from [45], and were used to compute local gradients and from this the slope (Supplementary Figs. S3-S5).

Electricity access estimates were calculated using visible nighttime satellite imagery from the SNPP-VIIRS data, downloaded from the Earth Observation Group at the Colorado School of Mines with a resolution of 500m [46]. The data are coded from 0-255 based on the brightness of measurable light with 0 being no measurable light in a location and 255 being scaled as the brightest level of measurable light. Population data from the LandScan database [47] was used as a measure of the ambient population (population over 24 hours) at a resolution of 30 arc seconds. Existing work has used visible nighttime light measurements from satellites as a measure for whether electrification is present [48]. Past work has shown this to be a reasonable measure, although the threshold sensitivity of the satellites results in listing some low population regions as no-light even if electrification may exist [49], [50]. For this study, lack of electricity access is defined as any location where population exists but no measurable nighttime light is visible. Further, to mitigate the low-population regions threshold limitation of nighttime lights data, only the population living in countries reported as having an incomplete electrification rate is considered [51]. This analysis was performed at 30 arc second (approximately 1 km$^2$) resolution globally. The geographic siting feasibility was assessed for every location on the globe at 3 arc second resolution (approximately 0.01 km$^2$). Locations were excluded that had no people living with no electricity access or in locations where land use restrictions cannot provide a contiguous area due to the presence of protected areas, wetlands and/or a land slope exceeding 3% (a conservative constraint for utility-scale photovoltaic arrays [52]). Geospatial (GIS) analysis was used to determine locations where feasibility of siting and available resources made solar-battery electricity development possible, Figure 1.

| Parameter | Unit of Measurement | Year | Resolution (Original Data) | Resolution (Study level) |
|---|---|---|---|---|
| Solar Irradiance [53] | watts per square meter | 25-year average (1983-2008) | 0.5º x 0.5º | 0.5º x 0.5º |
| Land Slope [45] | Elevation (m) | 2008 | 3 arc seconds | 3 arc seconds |
| Wetlands [54] | Binary: Wetland or not | 2004 | 30 arc seconds | 3 arc seconds (up-sampled) |
| Protected areas [44] | Binary: Protected area or not | 2015 | 30 arc seconds | 3 arc seconds (up-sampled) |

| Nighttime light values [55] | Unitless, measured from 0-255 | 2021 | 500 m | 30 arc seconds |
| Ambient Population [47] | number of persons per square kilometer | 2021 | 30 arc seconds | 30 arc seconds |

**Table 1.** Summary of data sources, resolution of original data and resolution of analysis.

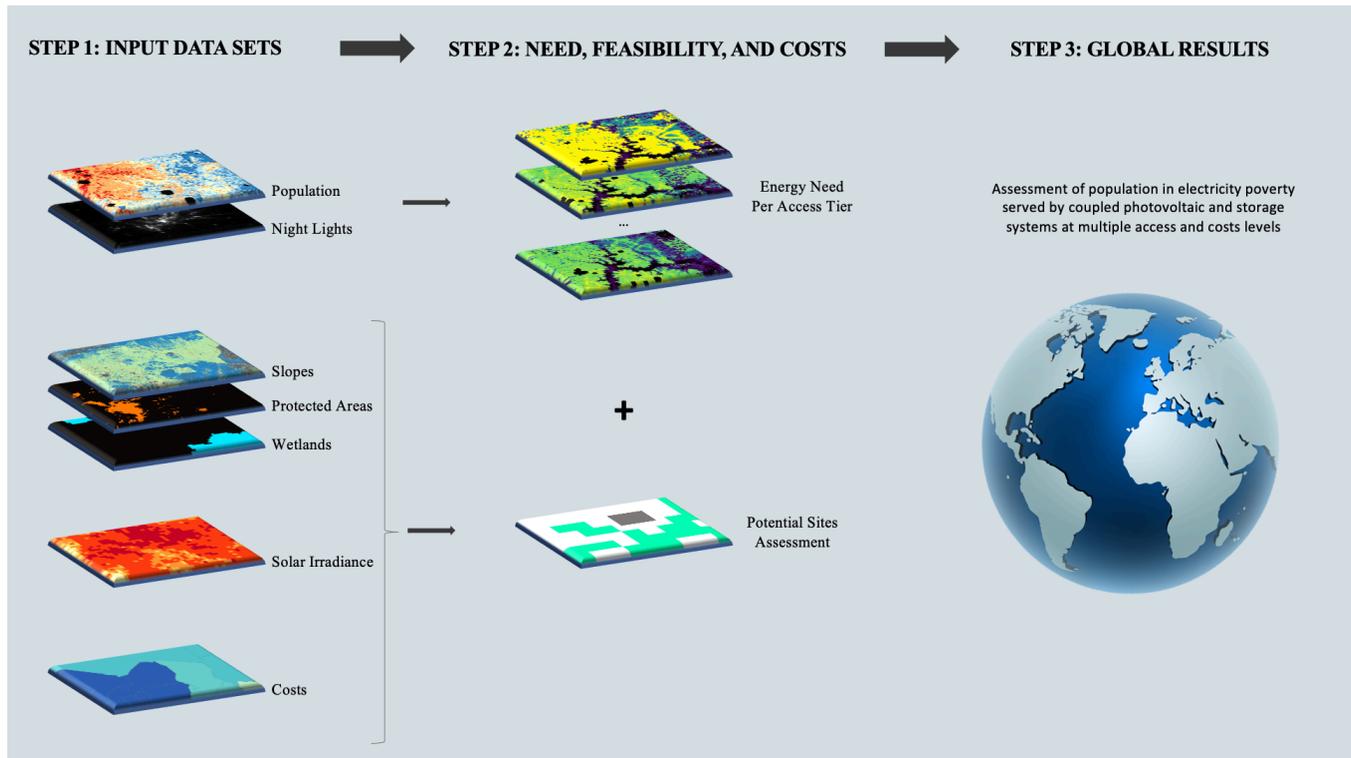

**Figure 1. Analysis process for combining multiple geospatial data sets to determine lack of electricity access, land use, natural resources, and people that can be served within each region at multiple access and costs levels.** Multiple geospatial datasets with different spatial resolutions were combined. Land-use layers were analyzed at up to 3 arc second resolution. Visible nighttime light and population layers were analyzed at 30 arc second resolution. Solar irradiance was analyzed at a 0.5° resolution. Land use restrictions and population and nighttime light data are downsampled within each 0.5° x 0.5° region to determine the population that can be served. The number of persons served is determined for each location where suitable resources exist to meet the minimum demand of household, productive and community uses for each Tier, and the costs are calculated from the optimal system found.

*Electricity Demand Estimation by Tier and Optimization of Energy Flow*
Coupled photovoltaic and battery storage systems were sized in every location to produce electricity generation profiles that met energy needs at every ESMAP Tier of access for the population in that location. The photovoltaic array and battery storage sizes were determined using an energy flow model that combined electricity demand, solar irradiance and physical system parameters to produce generation profiles. These profiles were compared with the demand profiles to determine capacity factors. Because a given level of access could be achieved by a range of photovoltaic array and battery sizes, the combination of array and battery size was chosen by optimizing for the lowest system capital cost. The optimization was done in every location and Tier of access over 25 years assuming 5% and 7% discount rates. Access levels in Tiers 3-5 were assumed to be met using utility-scale systems (including micro-, mini- and larger coupled storage and photovoltaic systems). Together, this led to 12 optimized energy flow dispatch scenarios in every location considered globally. In each scenario, the levelized cost of electricity was calculated based on array and battery size requirements and operations to meet at least the minimum criteria defined for each Tier throughout the entire 25-year analysis period.

The size of the array and battery were chosen to minimize the capital cost of the system. This was done assuming $1/Wp (Tiers 3-5) for the installed cost of the array. Battery systems costs were projected to 2030 using an exponential learning curve to historic lithium-ion battery prices [57], [58]. The 25-year dataset of global horizontal solar irradiance values used for this study were derived using an algorithm that has been critically evaluated and found to produce good results relative to ground-based measurements [59]. The D1 and DX datasets from the International Satellite Cloud and Climatology Project provided input data for the period from July 1983 through June 2008, with a 3 hour temporal resolution, and $0.5°$ x $0.5°$ spatial resolution [59]. Full details in Supplementary Note 2.

Electricity demand was estimated per capita using data for *household*, *productive* and *community* uses as prescribed in the Tiered framework as well as community *streetlighting* [21]. Amount and availability factors were calculated for each Tier using guidelines on the available electricity for different applications from [21]. For each of the three use categories, a different guideline is given (values per household, values for productive household members, and total value per household, respectively). Streetlight guidelines for communities is provided with no specific population basis. Therefore, estimates that assumed the most conservative guideline available were considered. Minute-level demand profiles for a representative household for one year for each tier from Narayan et al. [39] were used to create daily representative profiles for each month. These were then combined with estimates for streetlighting and productive and community uses. For streetlighting, per capita estimates for nighttime use were calculated (see Supplementary Note 1 for full details). For productive and community uses, minimum value estimates were taken from [21] and normalized over the day. This baseload is consistent with a range of estimates of commercial and industrial electricity profiles from a range of geographies and seasons. While there is some variation, the estimate used generally sees more consistent load than household-specific curves (e.g. [40], [41], [60]–[63]). The estimated daily values for productive, community, and streetlighting demand were added to the household Tier-specific curves used as inputs for system sizing analysis for each scenario in every location. Full details on demand profiles generated can be found in Supplementary Note 1.

**Results and discussion**. The 2030 Agenda for Sustainable Development highlights the clear ties between energy access, poverty reduction, and need for a low-carbon growth pattern [64]. The ambitious target of energy access for all [11] requires new ways of providing reliable, low-carbon energy. This analysis produces three results: land use and siting availability at a global level, dispatchable photovoltaic electrification potential using battery storage, and the resulting costs (LCOE). Results are presented for populations living with no nighttime light and within countries presenting an incomplete electrification rate.

Figure 2A shows the distribution of persons living in regions with no measurable nighttime light and within countries with an incomplete assessed electrification rate, which these results place at 1.15 billion globally. This is within the estimates of 1.0 to 1.6 billion people living without access to electricity [6], [7]. Figure 2B shows the capacity needed to meet the need of the local population (0.5-degree resolution) to a Tier 5 level. Sub-Saharan Africa and South-East Asia present the largest electricity deficit. While many regions of the world can be served with small systems (less than 10 MWe), many others require larger systems (more than 200 MWe). These systems may see economies of scale in construction and storage.

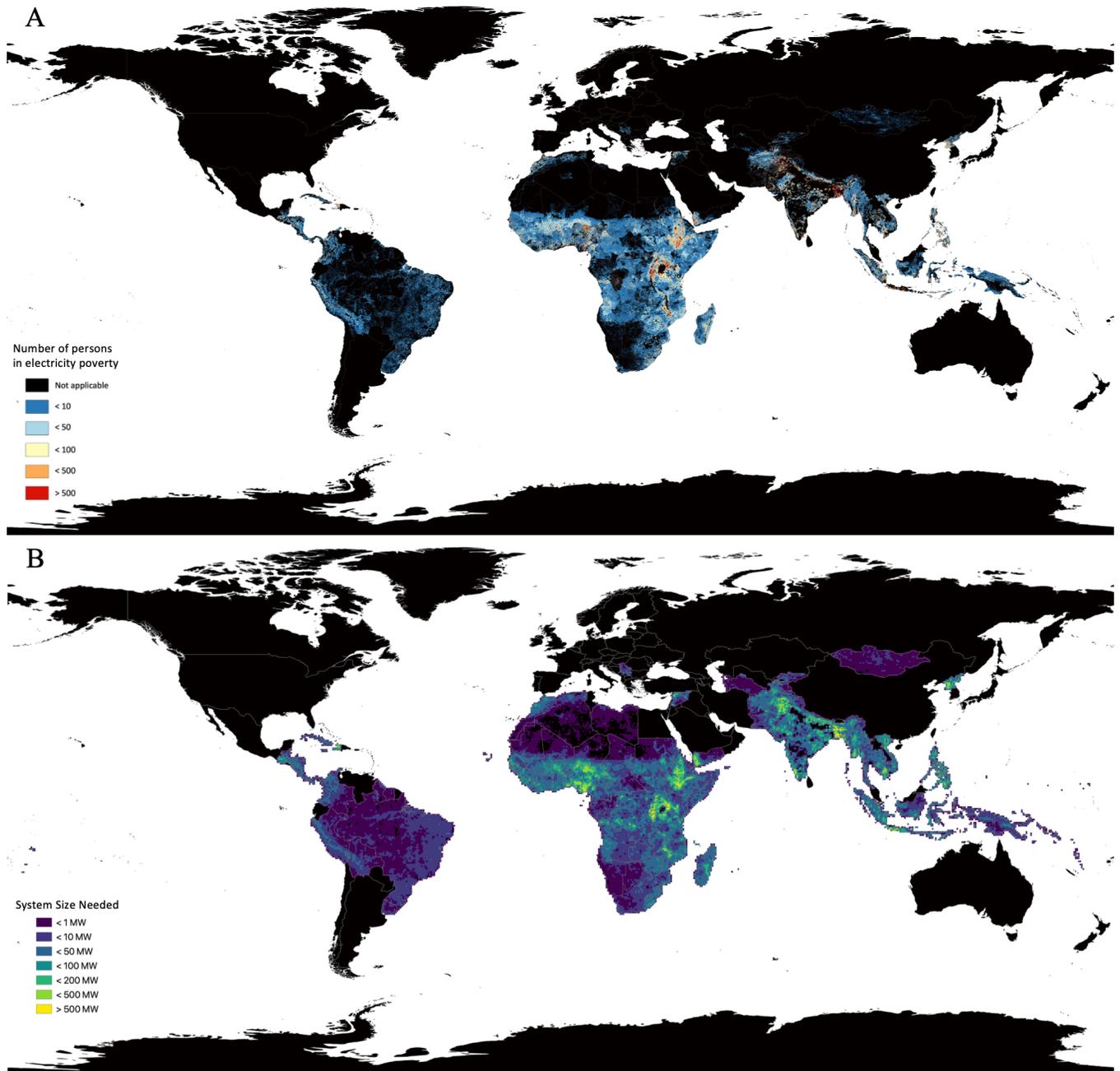

**Figure 2. (A) Total population living in areas with no measurable nighttime light within countries with an incomplete electrification.** The map shows the total population in each 30 arc second region where satellites measured no visible nighttime light. The most concentrated no light populations can be found in Sub-Saharan Africa, Central and South America, and Southeast Asia. Findings are consistent with other studies using similar methodologies [49], [50], [65]–[67]. (B) **Systems required to meet electricity needs of no-light population in each 0.5° x 0.5° region at Tier 5 level, for the countries presenting an incomplete electrification rate.** The results indicate the system size, in

megawatts-peak, required to serve the entire no-light population living in each 0.5º x 0.5º region at the Tier 5 level (0.95 capacity factor).

Figure 3 shows the fraction of land available (Fig 3A) and necessary (Fig 3B) for siting solar systems to serve the local population in each 0.5x0.5º regions based on land use parameters and identified need. Most locations present a suitable siting area for above 10% of the region (Figure 3A), while Figure 3B demonstrates that in most locations of interest to increase access to electricity, less than 1% of the land is necessary (Sub-Saharan Africa for example). More densely populated areas (India, South-East Asia) can see higher local area need, and some localized areas are found to not have enough suitable land to meet the demand (e.g., some areas of Bangladesh).

A

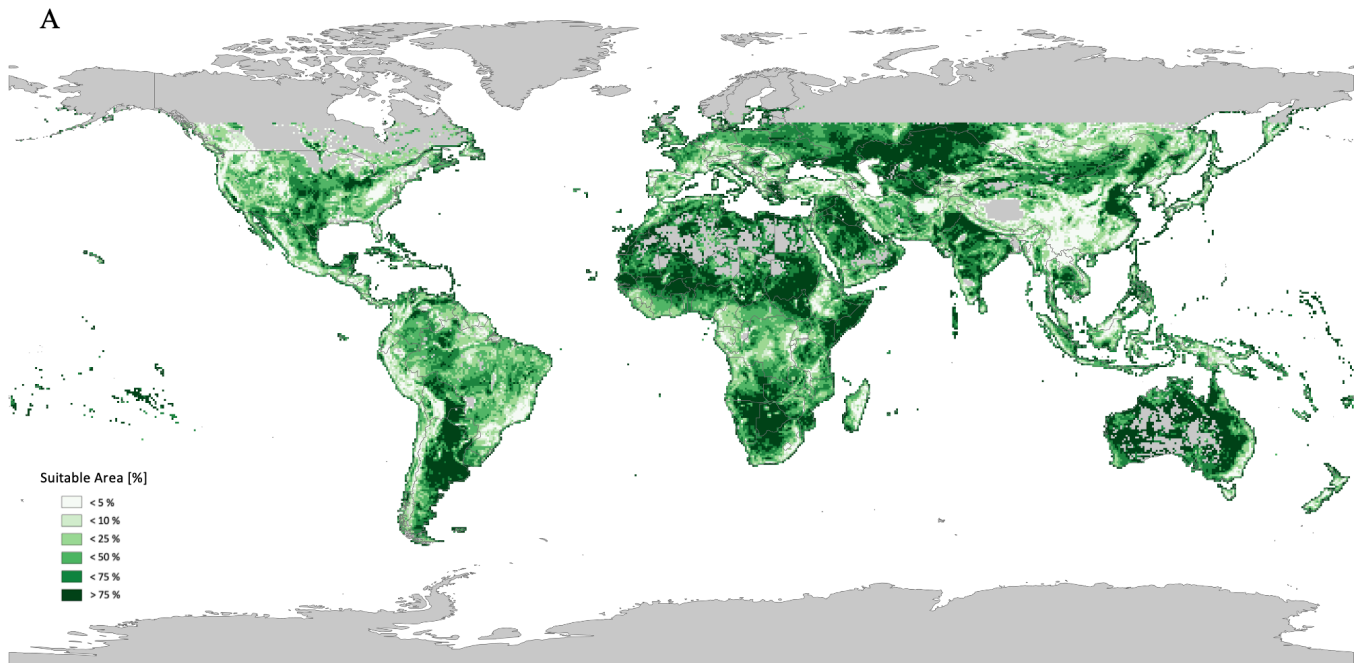

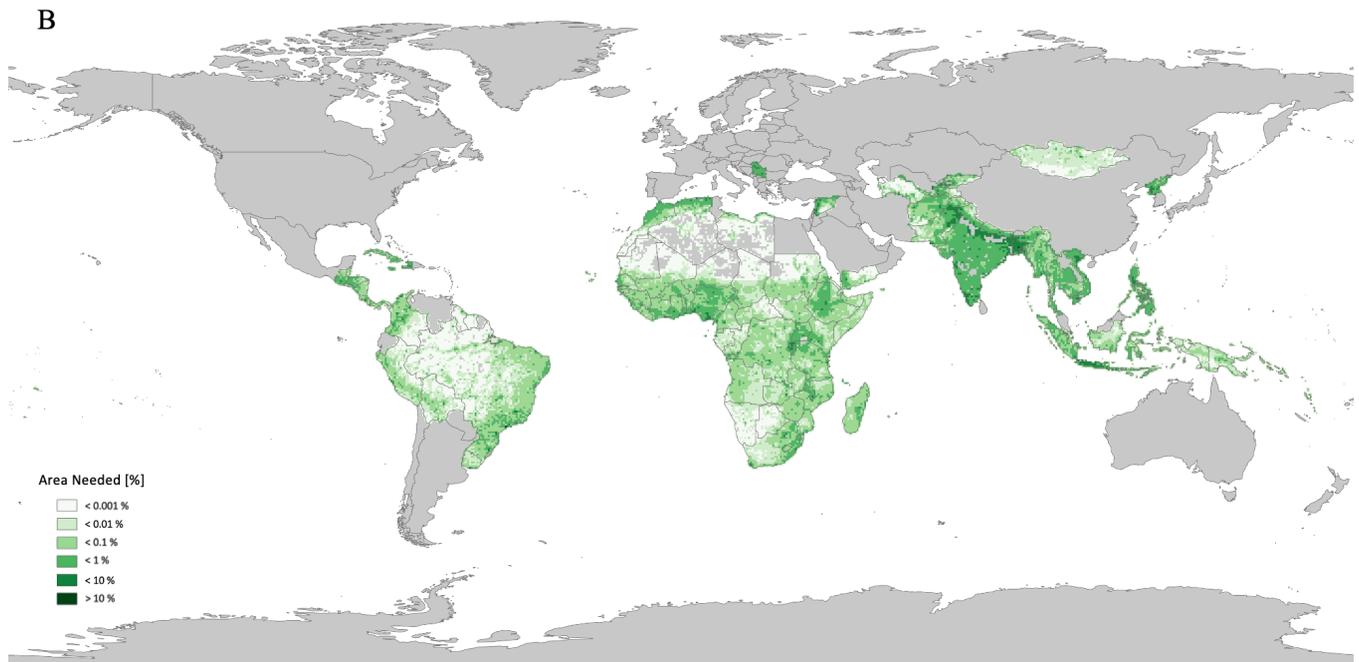

**Figure 3. [A] Distribution of (non-contiguous) land suitable for siting a photovoltaic array in each 0.5º x 0.5º geographic region.** The figure displays the results of the land use analysis performed at a 3 arc second resolution. Land is considered suitable if the area has a slope of less than 3 degrees and is not a protected area or a wetland (see Supplementary Note 2). Above a latitude of 60 degrees North, no analysis was completed because solar irradiance values fluctuate seasonally, and limit viability of photovoltaic systems coupled to storage for baseload systems. **[B] Fraction of land needed in each region to address the corresponding electricity needs.** The results indicate the fraction of land that must be available to serve the entire no-light population living in each 0.5º x 0.5º region at the Tier 5 level (0.95 capacity factor), within countries presenting an incomplete electrification rate.

Figure 4 shows the levelized cost of electricity for the number of persons served based on multiple cost scenarios at Tier 5 level. The discount rate and the cost scenario have substantial impacts on the number of people that can be served at or under a specific lifetime levelized cost of electricity. The levelized cost is dominated by the initial capital (construction) costs, resulting in higher lifetime costs for higher discount rates. Tier 5 has the most stringent requirements for service levels, including a reliability value (capacity factor) of 0.95, which requires high levels of battery storage for prolonged periods of cloudy weather and increases costs in many locations. Figure 4 shows that more than 90% of the population identified living without access to electricity (representing a billion people) could be served to Tier 5 level (8.2 kWh per capita per day) at less than 20 cents a kWhe using near-future cost estimates and for any considered discounting rate. Some scenarios, such as costs estimates by 2030 coupled with a discounting rate of 5%, even show a consequential electricity access increase potential for sub-10 cents a kWhe. Other tiered access levels and corresponding scenarios are shown in Supplementary Information 4 (Fig. S6 and S7) and exhibit slightly cheaper potentials.

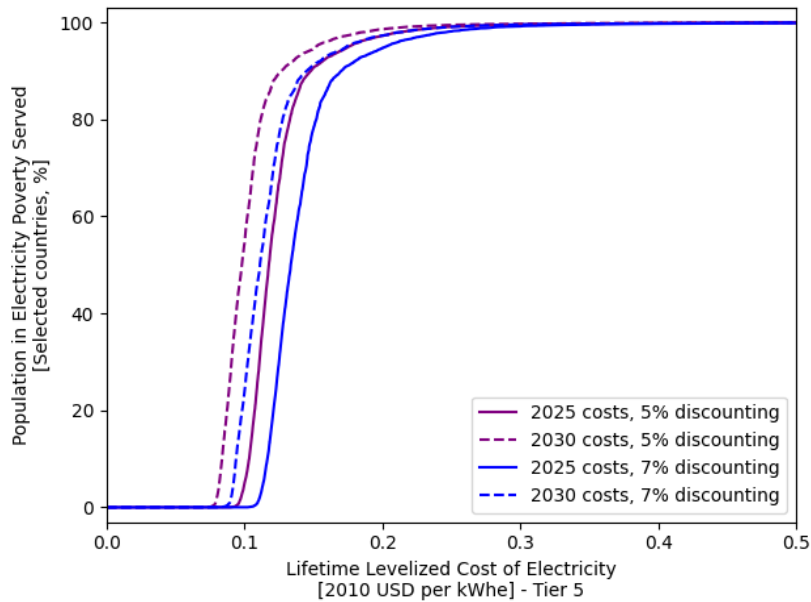

**Figure 4. Fraction of persons identified living without electricity access served at each value of levelized cost of electricity (LCOE) for Tier 5 access to electricity.** Multiple system costs scenarios from historical data and projected estimates are combined with a discounting rate at 5% and 7%.

Figure 5 shows a levelized costs of electricity map for Tier 5 access, assuming a 5% discounting rate and using future projected prices (2030 estimates). Projected costs improvement for solar and battery systems causes a drastic effect, with most of the electricity impoverished world showing a potential LCOE below 0.15 $/kWhe. While the costs vary geographically (and may vary additionally for grid distribution based on town density, supplies, labor and other factors), the ability to provide consistent power at a Tier 5 level (0.95 capacity factor) indicates that low-carbon photovoltaic-battery system deployment is feasible at most locations in need. This is especially notable in locations in Sub-Saharan Africa, where very high concentrations of persons living without electricity access are located in dry, remote regions (such as along the Sahel).

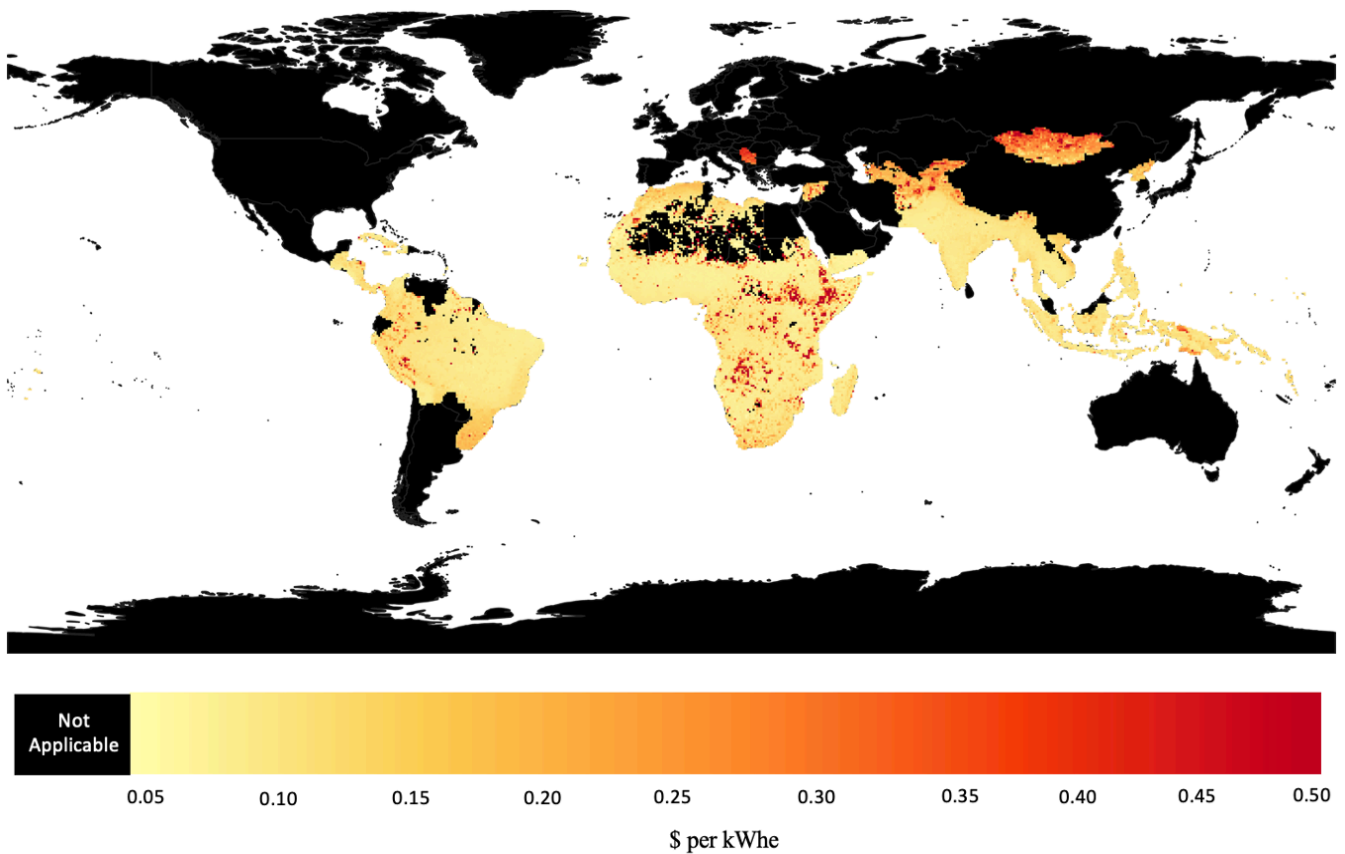

**Figure 5. Map of Levelized Cost of Electricity with future cost estimates (2030).** Shown for a Tier 5 electrification access targeted at population identified as living without electricity access, with a discounting rate of 5%. The data is given at a 0.5-degree resolution, in 2010 USD.

This study presents several limitations. While the use of nighttime lights combined with ambient population has been shown to be a good indicator of lack of electricity access, some regions may have sufficient access for their needs, yet be too rural to be detected by satellites. Additionally, we find that the system footprint on land use is below the physically feasible areas in most regions. However, competing land uses such as agriculture could have an impact on siting locally. Considering potential natural hazards impacts on system resilience, as well as climate change effect on energy needs, would also be of interest. This study considers homogeneous social environment, but local variations in security risks, difficult business environments, lack of serious providers or local materials, and other factors, would play a role in addition to the LCOE. Finally, the costs of solar system and batteries are rapidly evolving and may cause the results presented here to be an upper bound of population served under a given LCOE in the near future. At the same time, the development of the industry at such a scale could have an impact on supply chains and material markets.

**Conclusions.** The results presented identify geographic regions where coupled systems of photovoltaics and energy storage would be capable of providing dispatchable electricity to populations currently living with no or very low access. Nearly all the population identified without electricity access (approx. 1.1 billion people) could get access to Tier 5 level electricity in the Sustainable Energy for All initiative framework using photovoltaics and battery storage coupled systems. While the land use impacts are shown to be relatively small for the scale of the systems analyzed, consideration would still need to be given to competing uses such as agriculture. Localized phenomena of importance, such as flood plain activity or migratory pathways could also play an important role in deciding whether a specific location is suitable for siting technology and is an area for future work. Concerns about electricity reliability, resilience, and carbon signature should be inherent in considerations of expanding global energy access, but geography, resources and population need are also important factors for considering infrastructure investments at any location. Given existing global initiatives to expand electricity access, this study provides an important step in identifying where such technologies should be considered. The scale and geographical distribution of the systems would minimize the required transmission infrastructure over large distances, reduce individual capital investments, and minimize disruption potential from hazards.

**Acknowledgements**.  We thank Christopher Elvidge and Tilottama Ghosh for discussions about the use of data on nighttime and population, Morgan Bazilian for discussion on lack of electricity access and development, and the US Department of Energy for funding under award 16-10730.

# Supplementary Information
# Potential of Photovoltaics and Energy Storage to Address Lack of Electricity Access

GF L'Her, AG Osborne, AE Schweikert, CS Ramstein, BL Stoll, MR Deinert

**Supplementary Note 1.** The choice of electrification standards (e.g., per capita consumption, capacity factor, home generation -vs- grid/microgrid access) is a challenge because it can be subjective. However, the Sustainable Energy for All (SE4All) program has established five Tiers of electricity access. These range from no access (Tier 0) to Tier 5 designation which can be considered approximately equivalent to electrification in many middle-income countries (as defined by the World Bank). The clear objective is to promote electrification from basic access (Tier 1) and move towards the Tier 5 category. Each of the Sustainable Energy for All electrification Tiers has multiple criteria which include per capita provision, capacity factor, community and productive uses and means of supply (e.g., through a grid or not).

In this analysis, all Tiers in the SE4All framework are considered. Narayan et al.[1] gives 1-minute resolution load profiles for all Tiers for one year. From this data, a representative daily profile for each month was generated. Minute-level profiles were compared to hourly average profiles. The hourly profiles showed only a slight smoothing of peak demand when hourly averages were used compared to minute-level resolution data. The hourly-level data matches the temporal resolution of the solar irradiance data used in this study for estimating generation profiles. Thus, the final demand profiles used for this analysis were a representative day for each month in the year at hourly-level resolution, Fig. S1. These profiles are representative for Sub-Saharan Africa. The majority of persons living in electricity poverty are located in the Global South, with rural Africa electrified at rates of only 25%, while Sub-Saharan Africa in total is estimated to have an electrification rate of 43%. This compares to a global access rate of 87%[2]. It should be noted that these profiles have limited seasonal variation, due to their tropical geography. While not all locations with no visible nighttime light will follow exact demand patterns, the vast majority of populations living in electricity poverty live in similar geographies (Central and South America, Southeast Asia). Figure S1 details the load profiles for each Tier in January and June to illustrate the temporal changes for each Tier.

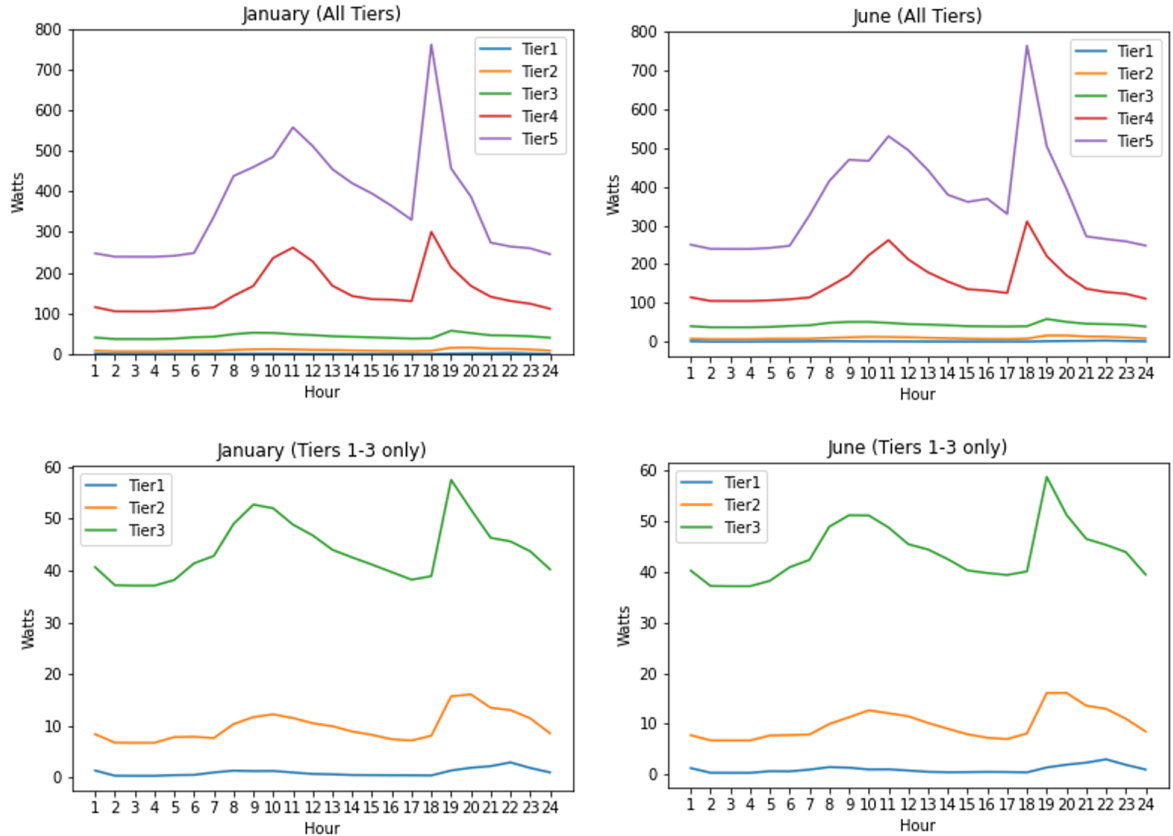

**Figure S1: Hourly load profiles for January and June. Profiles assembled as a representative day in each month, based on minute-level data from Narayan et al. [1]. A representative profile for every month was generated for this analysis. The temporal resolution matches available historical solar insolation data.**

In addition to household level data, the load for commercial, productive and street lighting uses was estimated to match all criteria of the five Tier framework. The per-capita load for streetlighting varies based on urban vs. rural environments, driving vs. walking roads and communities, environmental considerations (light pollution) and many other factors. Estimates range greatly, with 1 light per 28 persons in rural Africa [3] and 1 light per 12 persons on average in the United States [4]. Recent investments made by the Indian government suggest approximately 1 light per 7-10 persons in urban areas [5], [6]. These estimates were aggregated into a conservative estimate of 1 light per 10 persons in this analysis. The amount of electricity required per light per night was estimated based on assumption of LED overhead streetlight installation. A midpoint of brightness/lumens was chosen between 50-260 watts (4,000-22,000 lumens) using available data cite ranges of 18-91kWh consumption per month [7]. The average per capita streetlight daily value was then adjusted based on community coverage factors provided in [8], ranging from 1% (Tier 1) to 95% (Tier 5). Full inputs used in this assessment are listed in Table SI.

**Table SI: System performance and availability summary of suitable electricity provision per tier used for analysis.** Values are minimum daily per capita estimates, based on guidelines found in [8]. Where data were given in hours per day, estimates were made based on 12-hour day/night daily pattern. Values include household, productive, community uses and streetlighting per capita estimates. Productive and community use is normalized as a baseload demand throughout each day, as is streetlighting. Where multiple values (*small* vs. *large* facilities for Community Infrastructure, for example) are given, the larger demand was assumed.

|  |  | Daily Watts [Watts] | Watt Hours [kWh] | Reliability Values (Minimum) | Nighttime Availability (Minimum) |
|---|---|---|---|---|---|
| *Household*, *Productive Applications,* and *Community Infrastructure* Demand Parameters | Tier 1 | 3 | 0.012 | 17% 8.3% | 8% |
|  | Tier 2 | 50 | 0.2 | 17% | 16% |
|  | Tier 3 | 200 | 1 | 33% 50% | 25% |
|  | Tier 4 | 800 | 3.4 | 75% | No minimum |
|  | Tier 5 | 2000 | 8.2 | 95% | No minimum |
|  |  | Daily Watt Hours [kWh per capita] |  | Nighttime Availability (Minimum) |  |
| Streetlighting demand estimates | Tier 1 | 0.115 |  | 17% |  |
|  | Tier 2 |  |  | 33% |  |
|  | Tier 3 |  |  | 50% |  |
|  | Tier 4 |  |  | 75% |  |
|  | Tier 5 |  |  | 95% |  |

**Supplementary Note 2**. The spatial resolution of the analysis in this work is 0.5°×0.5° globally. Population and visible nighttime light data (Figure S6 and S7) are available at 30 arc-seconds resolution. To determine population demand, the total population (all persons) and population living in no light regions is aggregated to the 0.5° (1800 arc-second) grid space. Therefore, the total energy needed, system sizing and other components are measured globally for every 0.5° cell.

The total energy output of a photovoltaic system in a given location, $E_i$ [kWhe], was divided into three components, Fig. S2:

$$E_j = E_g + E_s + E_e \qquad (1).$$

Here $E_g$ [kWhe] is the energy sent directly to the local microgrid to meet electricity needs and $E_s$ [kWhe] is the energy directed into storage when the output of the array exceeds immediate needs. The systems are sized to produce electricity in each grid cell with a 25-year minimum capacity factor given by the ESMAP Tiers 1-5, 8.33%, 16.67%, 50%, 75%, 95%, respectively. Five load curves are considered that correspond to the ESMAP Tiers, Fig. S1. When the storage is full and the array produces more than load curve demand, $D$ [kWhe], the excess electricity $E_e$ [kWhe] is also available for broader use or is assumed to be curtailed, Figure S2. Energy is provided from storage, $E_{fs}$ [kWhe] during the night and during the day when the output of the array is below load curve demand. The capacity factor was then defined as the fraction of hours for which the energy sent to the grid by the plant met or exceeded demand:

$$CF = \frac{\sum_{hours}(E_g + E_{fs} \geq D)}{25 \text{ years} \times 8760 \text{ hours per year}} \qquad (2),$$

where the sum is over the number of hours in horizontal global irradiance dataset (described below).

The build size of industrial scale PV facilities -vs- the panel size has been well studied and is typically found to be ~32% greater than the panel size for facilities > 20MWe and ~38% bigger for systems < 20Mwe [9]. We chose the upper bound of this range (i.e., 38%) to be conservative. The difference between panel area and total facility area is considered in this analysis.

A 25-year data set of horizontal global irradiance with 0.5°×0.5° spatial, and 1-hour temporal, resolution was used as a basis to

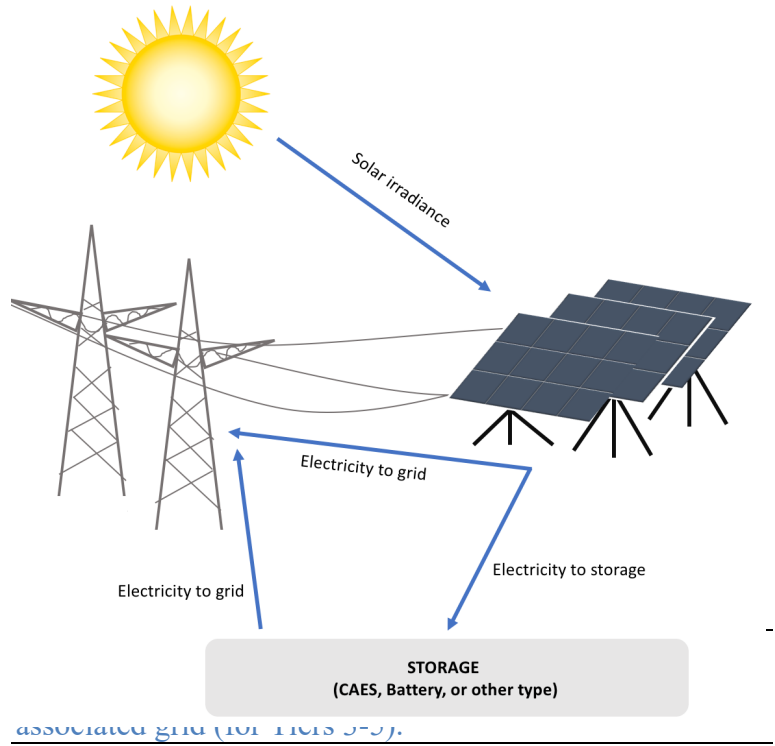

associated grid (for Tiers 3-5).

compute time-dependent electricity generation profiles. The size of the array and storage system were chosen to minimize the capital cost of the system assuming $1/Wp for the installed cost of the array for Tiers 3-5, and the battery storage costs given in Table SII. Additional input costs for Tiers 1 through 5 are also given in Table SII. Electricity produced is determined for each 0.5°×0.5° region based on demand, array sizing, solar irradiance and battery storage. The sizing of photovoltaic array and battery is based on the population in each 0.5°×0.5° region, the corresponding load profile, and the level of access corresponding to each Tier.

Global solar irradiance time series data for each 0.5°×0.5° location are given in Universal Coordinated Time, while load curve data correspond to local time. Time synchronization of global solar irradiance values with local load curves was done by shifting the local load curve according to each location's time zone offset. Global time zone data were taken from [10] which compiled original IANA [11] time zone codes from OpenStreetMap into shapefile format. The time zone codes were projected onto the 0.5°×0.5° grid and time offsets from Universal Coordinated Time were computed by combining time zone codes with built-in Matlab functions.

For access Tiers 3-5, input capital costs for batteries are on a CAPEX basis, which includes financing during construction. Battery costs for Tier 3-5 levels of access in 2025 and 2030 were computed by fitting an exponential learning curve to historic lithium-ion battery prices in 2010-11 and 2018 [12], [13]. Our costing model accounts for fixed and variable operating costs, and capital costs separated into AC (power) and DC (storage capacity) components. The AC system's cost is incurred according to the peak discharge rate of the battery, which is set equal to the peak power capacity of the photovoltaic array. The degradation of battery storage capacity as a function of time is computed using a depth-of-discharge model [14]. The depth-of-discharge at a location used to calculate degradation is computed by taking the average depth-of-discharge in every charge-discharge cycle over 25 years. Degradation is calculated annually and replaced on the same annual basis, the costs for which are computed by discounting the CAPEX cost of storage capacity replaced at the time of replacement.

Table SII lists cost inputs used in the analysis. In the computation, costs are discounted to 2010 US dollars using a GDP deflator from the Bureau of Economic Analysis [15].

**Table SII. Inputs to the pricing model (2018 USD), Mongird et al. (2019) [13]**

| COMPONENT | 2025 PROJECTED COST | 2030 PROJECTED COST |
|---|---|---|
| RESERVOIR COST (GRID-SCALE) | $177.72/kWhe | $78.32/kWhe |
| GENERATION (POWER) COST (GRID-SCALE) | $185.73/kW | $113.08/kW |

**Supplementary Note 3**. Data used to determine suitable land use for siting a photovoltaic array with coupled storage are shown in Figures S3 – S5.

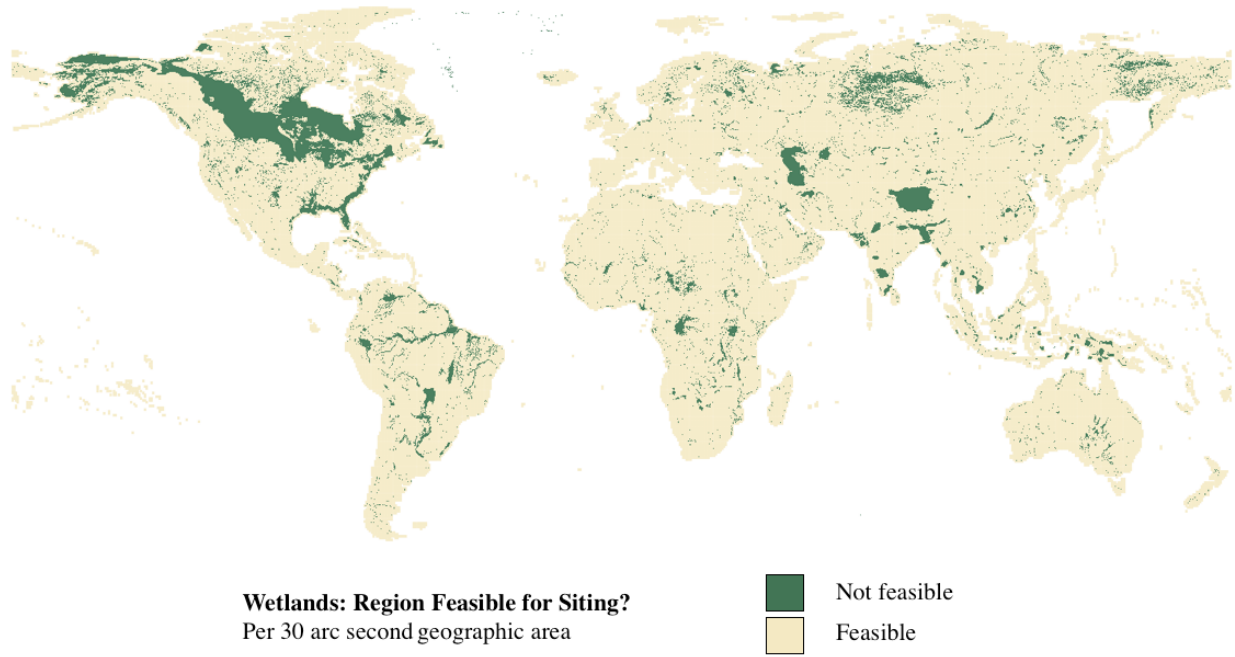

**Figure S3. Global Wetlands**. Wetland regions with a surface area ≥ 0.1km² are shown in green. Data on the distribution of global wetlands was taken from [16]. These data include all lakes, reservoirs and wetlands with a surface area of ≥ 0.1km². All types of wetland data were included in this analysis and excluded as possible locations for siting a photovoltaic array.

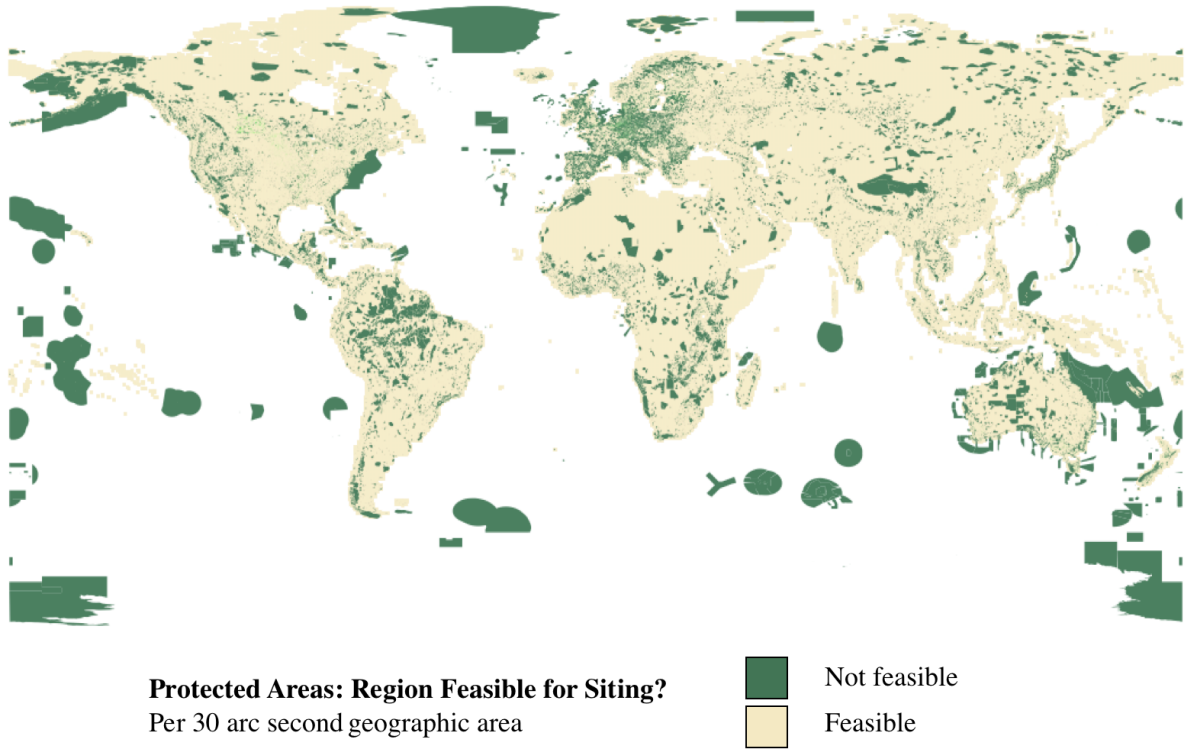

**Figure S4. Global Protected Areas.** Protected regions are shown in green. Data on protected areas from [17]. These areas include both marine and terrestrial areas and are excluded from possible locations for siting a photovoltaic array.

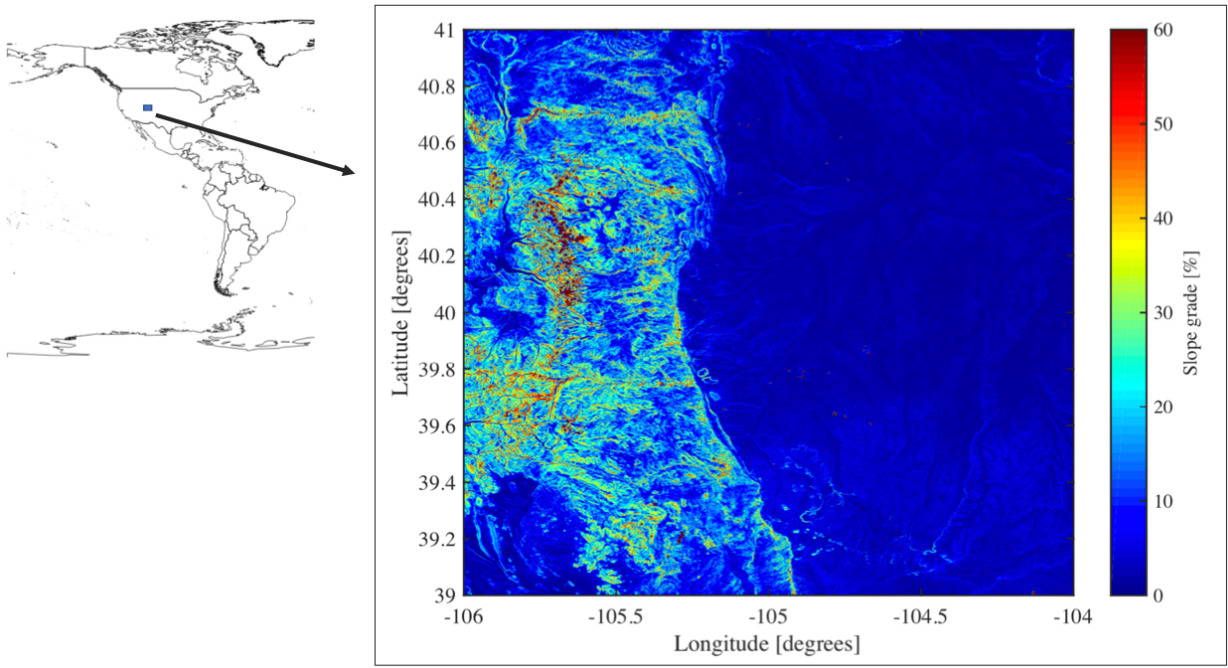

**Figure S5. Land Slope.** Data on land elevation with 3 arc second resolution were used for this study to determine land slope. We show a section of land in North America at approximately 40 degrees Latitude and 105 degrees Longitude.

**Supplementary Note 4**. Figures S6 and S7 show the fraction of the population identified as living in electricity poverty that could be served by a coupled photovoltaic and energy storage system for different access tiers and for different initial costs and discounting rates scenarios, at multiple LCOE levels.

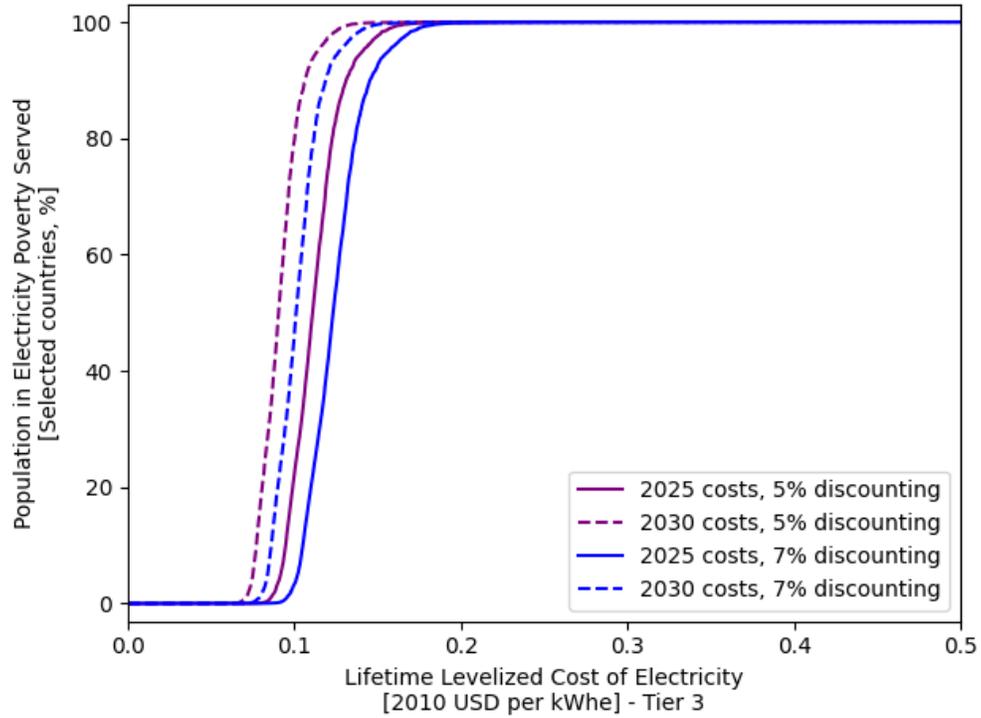

**Figure S6. Fraction of population in electricity poverty served by a coupled photovoltaic-battery system under different scenario assumptions.** Tier 3 represents an access level of 1 kWhe per capita and a reliability of 50%.

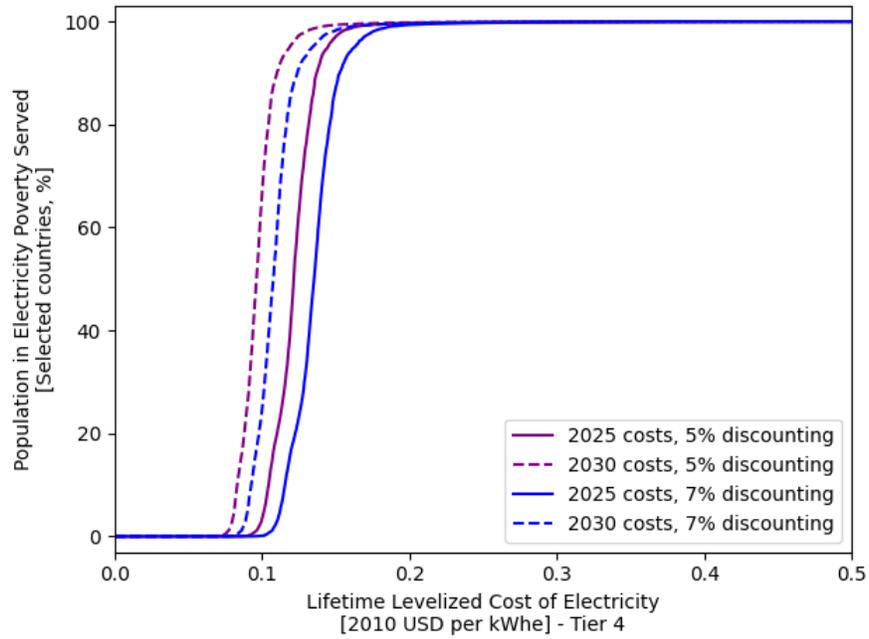

**Figure S7. Fraction of population in electricity poverty served by a coupled photovoltaic-battery system under different scenario assumptions.** Tier 4 represents an access level of 3.4 kWhe per capita and a reliability of 75%.